# Spontaneous emission of color centers at 4eV in hexagonal boron nitride under hydrostatic pressure

*Kamil Koronski[1], Agata Kaminska[1,2], Nikolai D. Zhigadlo[3,4], Christine Elias[5], Guillaume Cassabois[5] and Bernard Gil[5,*]*

1-Institute of Physics, Polish Academy of Sciences, Al. Lotnikow 32/46, PL-02-668 Warsaw, Poland

2-Cardinal Stefan Wyszynski University, Faculty of Mathematics and Natural Sciences. School of Exact Sciences, Dewajtis 5, PL-01815 Warsaw, Poland

3-Department of Chemistry and Biochemistry, University of Bern, 3012 Bern, Switzerland

4-CrysMat Company, 8046 Zurich, Switzerland

5-Laboratoire Charles Coulomb, UMR 5221, CNRS-Université de Montpellier, 34095 Montpellier, France

* bernard.gil@umontpellier.fr

**Abstract:** *The light emission properties of color centers emitting in 3.3-4 eV region are investigated for hydrostatic pressures ranging up to 5GPa at liquid helium temperature. The light emission energy decreases with pressure less sensitively than the bandgap. This behavior at variance from the shift of the bandgap is typical of deep traps. Interestingly, hydrostatic pressure reveals the existence of levels that vary differently under pressure (smaller increase of the emission wavelength compared to the rest of the levels in this energy region or even decrease of it) with pressure. This discovery enriches the physics of the color centers operating in the UV in hBN.*

## I-    INTRODUCTION

Boron nitride is a wide bandgap semiconductor which was synthetized as early as 1842 by Balmain[1]. Many chemical methods have been since developed to synthesize hBN powders, bulk crystals, nanosheets, and nanotubes for different applications. It took more than one century and a half until high quality, single crystals with macroscopic millimetric size were produced[2] leading to realization of a light-emitting device operating in the deep UV[3]. This demonstrator paved the way for applications of hBN to advanced optoelectronics[4], making him to be considered as a challenger of aluminum nitride[5]. There is currently a tremendous activity for realizing efficient solid state Single Photon Sources (SPSs), based on color centers and susceptible to operate at room temperature. Several demonstrators were already published in case of operation in the neighborhood of the long-wavelength portion of the light spectrum[6-10]. Xue et al.[10] have in particular measured an anomalous pressure coefficient[11]. Bourrelier et al.[12] have reported single photon emission of color centers emitting at 4.1 eV. The question of the nature of the defect giving rise to this behavior is under debate, at the time being. We have decided to measure the evolution of this photoluminescence in the crystal submitted to a hydrostatic pressure to contribute to the elucidation of the origin of such emission.

## II- EXPERIMENTAL DETAILS

### II-1- Growth of high quality hBN under high pressure

The hBN crystals we use here were grown by means of a high-pressure, high- temperature (HPHT) cubic anvil technique, developed earlier for growing superconductors and various other compounds[13]. The details of experimental setup can be found in previous publications [14,15]. A mixture of Mg (99.99% pure), B (99.99%) and BN (99.9%) in a molar ratio 1:1.2:0.1 to give a total mass of 1 g were used as starting materials further enclosed in a BN crucible. Following Taniguchi and Watanabe[16], the BN sources were purified by heating them at 2000 °C for 2 h under dynamic vacuum in order to remove oxygen impurities and other volatile contaminants. The sample was compressed to the desired pressure (above 2.5GPa) at room temperature. Then, the temperature was ramped up at constant pressure to the desired value near or above 1800°C. It takes about 1 hour to reach this plateau which is kept and controlled during about another hour and then cooled down to 20 °C in 1 h. Afterwards, the pressure was released and the sample removed at the end of a full growth process lasting around 4 hours. After mechanically crushing the BN crucible, the high-pressure product was heated in vacuum at 750 °C for 0.5 h to remove the excess Mg. Although the Mg solvent apparently plays an important role, the crystal growth process of hBN is not a simple precipitation from a solution in molten magnesium. Rather, the hBN crystals are the product of a reaction in the ternary Mg–B–N system which allows the simultaneous crystal growth of completely different types of materials: a wide band gap semiconductor (hBN) and a superconductor ($MgB_2$). By varying the synthesis conditions, it can be determined the sequence of phase transformations occurring in the Mg–B–N system[13]. Raman spectroscopy measurements and X-ray diffraction experiments confirm the good quality of the grown crystals.

### II-2- Photoluminescence at low temperature and atmospheric pressure

In figure 1 is reported on a logarithmic scale the 8K photoluminescence (PL) spectrum recorded using as an excitation wavelength the fourth harmonic of a Ti:Sa laser taken on a sample which photograph is inserted in the figure. The excitation wavelength was 196.5 nm and the full set-up was described in Ref.[17]. The spectra were normalized relatively to the value of the PL intensity at 5.47 eV to show the spectral variation of the PL intensity over three decades in the range of investigated energies. All samples exhibit the classical TA(T), LA(T), TO(T), LO(T), phonon-assisted transitions at 5.89 eV, 5.86 eV, 5.79 eV and 5.76 eV, respectively. These transitions establish the indirect nature of the bandgap of hBN. Zooming at higher energy it is possible to detect the signature of the forbidden Z-type phonon assisted transition at 5.93 eV. At lower energy, i.e. in the 5.6 eV to 5.0 eV range, there is a series of lines equally spaced by 147 meV that is the energy of a TO(K) phonon. The recombination is ruled by the joint and subtle contributions of two different actors; namely an intervalley scattering process which contributes to the value of the matrix element for the different overtones[18], and defects first identified by Bourrelier et al.[19] which densities at the energies of intervalley scattering processes contribute to the density of state in the Fermi's golden rule. Slight fluctuations of the densities of the different defects may s modulate the overall shape of the PL of bulk crystal in the 5.6 eV to 5.32 eV energy range, from a sample to another one. It is worthwhile noticing that it is possible to get rid of such features in cathodoluminescence (CL) experiments using as a probe an electron beam smaller in size than our laser spot to shine defect-free regions of the crystal[20]. At 5.3 eV and 5.56 eV are detected two PL features that fail out of the two trends described above and which are interpreted in terms of the signatures of the nitrogen vacancy and an un-identified defect. It is important to recall that the intensity of the latter PL line is  far less robust with T than all the other lines[18].

Scaling to the low energy range, a weak PL hump is detected in the sample contaminated with magnesium, this is not significant and a stronger one is always recorded in the 4 eV range to which emerge a series of sharp lines, signature of a color center with a zero-phonon line at 4.1 eV and a set of phonon assisted lower energy companions that sample the whole Brillouin zone[21]. We note that this line increases in case of carbon doping/contamination, which was noticed by Taniguchi and Watanabe[16]. Their attribution to carbon and/or oxygen has been recently contested by Tsushima et al.[22] and by Chichibu et al.[23]. The latter attributes it to a carbon on the N site together with an oxygen on the N site, and the former attributes the origin of this PL line to the boron vacancy and refutes a correlation with the concentration of trace impurities like carbon and oxygen.

*The important message we infer from this PL spectrum is that we deal with a high quality BN sample, an important condition for high-pressure investigations.*

### II-3- Photoluminescence of hBN under high pressure

The high hydrostatic pressure was obtained by using a low-temperature diamond anvil cell (CryoDAC LT, easyLab Technologies Ltd). Argon was used as a pressure transmitting medium. The DAC was mounted in an Oxford Optistat CF cryostat equipped with a temperature controller for low temperature measurements. A piece of sample with a thickness of around 30 μm and diameter 100 μm was loaded into the cell along with a small ruby crystal. The $R_1$-line of ruby luminescence was used for pressure calibration. The line-width was also used for monitoring hydrostatic conditions in the DAC. The PL of  hBN was excited using deep-UV 275.4 nm laser line of a continuous wave (CW) Coherent Innova 400 Ar-ion laser. At this energy of 4.5 eV, diamond is transparent. This is a below-bandgap excitation but the number of photo-created electron and hole pairs in line with absorption of the laser to defects at 4.5 eV is high enough to permit us recording PL with a good signal to noise ratio. The emission was dispersed by a Horiba Jobin-Yvon FHR 1000 monochromator, and the signal was detected by a liquid nitrogen cooled charge-coupled device camera. The scattering of laser light leads to a technical difficulty to measure PL features at energies higher than 4 eV during our experiment.

In figure 2 are plotted several normalized PL features recorded in the 1GPa to 5 GPa range. We remark that most of the lines shift toward low energy when increasing pressure. This is interesting in particular as the line at 4.1 eV at atmospheric pressure can be progressively detected when increasing the pressure as can be seen in the right-hand side of the figure for pressures beyond 2.5 GPa. We can unambiguously record the most intense phonon replicas of this zero-phonon line which energy is extrapolated to its expected value at atmospheric pressure. When increasing the pressure, the PL bands broaden, probably due to in-homogeneities of the applied pressure (using an argon-based technology is not the best for working at liquid helium temperature), but the full width at half maximum of the zero-phonon line and of its two LO($\Gamma$) replicas behave in concert. It is worthwhile noticing the detection of two transitions below 3.4 eV under the excitation at this laser wavelength. The higher energy line is particularly strong at 1.5 GPa and its pressure coefficient is positive while the lower energy level red-shifts when increasing the pressure. There are thus at least two kinds of defects that contribute to the PL of hBN in the 3.3-4.1 eV range. The pressure changes of relative intensities of different the PL lines can be attributed to pressure shifts of excitation bands of different defect levels, influencing excitation efficiencies and PL intensities of related lines.

The evolution of the shape with pressure is suggestive of the existence of a third actor: a broad underlying band. Nothing in the measured behaviors ( red-shifts overlapping with blue-shifts) is specific to boron nitride; similar trends were already reported for other materials. We now discuss the numbers that are specifics of this investigation.

## III- DISCUSSION

### III-1 Pressure dependence of electron states and phonon states in hBN

According to group theory and to the physical properties of uniaxial crystals, an hydrostatic pressure P does not reduce the $D_{6h}$ symmetry of boron nitride[24,25].

Therefore, the modification of the energy $\delta E_\alpha^i$ of an extrema i of any conduction or valence band $\alpha$ under the application of an hydrostatic pressure $\delta P$ writes within the context of perturbations theory [26] in the linear approximation as:

$$\delta E_\alpha^i = \left[\left(\frac{\partial E_\alpha^i}{\partial a}\right)_c \frac{da}{dP} + \left(\frac{\partial E_\alpha^i}{\partial c}\right)_a \frac{dc}{dP}\right] \delta P + o(\delta P^2) \tag{1}$$

where a and c are the lattice parameters of boron nitride and $\frac{da}{dP}$ and $\frac{dc}{dP}$ represent their derivative under an hydrostatic pressure.

Thus, under pressure, the evolution of the energy difference between extrema $\alpha$ and $\beta$ of two bands i and j is simply given by the difference $\delta E_\alpha^i - \Delta\delta E_\beta^j$.

Energies of optical transitions are ruled by a very similar difference, to which however has to be added, in case of transitions indirect in the dual space, the contribution of a phonon with an ad-hoc wave vector $k_\alpha - k_\beta$ that fulfills the k-selection rule[27].

The energy $\hbar\omega_{k_\alpha - k_\beta}$ of such phonon also varies with pressure[28]:

$$\delta\hbar\omega_{k_\alpha - k_\beta} = \left[\left(\frac{\partial\hbar\omega_{k_\alpha - k_\beta}}{\partial a}\right)_c \frac{da}{dP} + \left(\frac{\partial\hbar\omega_{k_\alpha - k_\beta}}{\partial c}\right)_a \frac{dc}{dP}\right] \delta P \tag{2}$$

Obviously, any accurate calculation of the band structure of a periodic crystal and of its phonon energies, both through the whole Brillouin zone and against the values of its lattice parameters, can in principle give access to the pressure dependence of a given transition, let it be direct or indirect in reciprocal space.

This statement should be moderated taking into account potential pressure-induced modifications of the band curvatures and of the dielectric constant which may impact the interpretation of experimental data apart from the simple scenario described above[28]. Also, from the experimental point of view there are often some important issues that prevent to observe such subtle effects among which are the various consequences that an unavoidable departure from a perfect hydrostatic perturbation may have on the energy diagram or non-linear effects in moderately hard materials, or onset of a phase transition towards a more compact crystalline state[29,30].

In case of crystal imperfections giving birth to localized defects that break the translational symmetry of the crystal, the wave functions of electrons $\Psi(e)$ and holes $\Psi(h)$ expand over the whole bands as[31]:

$$\Psi(e,h) = \sum_k^i c_k^i(e,h) |\varphi_k^i(e,h)\rangle \qquad (3)$$

Where the $|\varphi_k^i>$ represent the Bloch waves for band $i$ at wave-vector $k$ and the $c_k^i$ are coefficients of the expansion correlated to the Fourier transform of the localizing potential:

$$c_k^i(u) = \langle \varphi_k^i(u)|\Psi(u)\rangle \qquad (4)$$

The evolution of the energy level under pressure is therefore given by a $c_k^i$ weighted average value of the pressure dependence of the different levels of the band structure that contribute to its wave function.

The influence of pressure on the properties of cubic (that is to say non uniaxial) crystals is very much documented in terms of the pressure-dependence of band, impurities and defect energies. Shallow hydrogenic levels (like for instance sulfur in gallium phosphide) which may be described within the context of the effective mass approximation do follow the pressure dependence of the critical point of the band structure nearest to them[32-34]. In some specific cases like typically for gallium arsenide, in the neighborhood of 4 GPa, a crossover occurs between the $\Gamma$ and $X$ states of the conduction band and a transition from direct to indirect is pressure induced for the fundamental bandgap. Then the blue-shift of photoluminescence energy (of about 107 meV/GPa) switches to a redshift at rate -14.5 meV/GPa[35]. Simultaneously, the photoluminescence intensity collapses in straightforward relationship with the reduction of radiative recombination rate when the indirect bandgap configuration is reached[36]. Such pressure-induced transition from direct toward indirect band gap structure is not specific of cubic semiconductors; it was also recently reported in $MoS_2$ monolayer[37]. The behaviors of a lot of materials have been also studied under pressure, including non cubic semiconductors, and in particular layered compounds like for instance InSe[38,39], or GaS[40] or transition metal dichalcogenides[41].

In case of hexagonal boron nitride, the shift of the indirect bandgap under hydrostatic pressure was measured once using a sapphire anvil cell for pressures up to 2.3GPa. This study revealed a redshift of the absorption edge at a rate of -36 ±1 meV/GPa[42]. By an extrapolation *mutatis mutandis* of the prediction of Hess and Dow[43] for cubic semiconductors, we imagine that this negative coefficient is indicative of an optical transition involving electronic states away from zone center, which is effectively true. Akamura et al.[44] measured by absorption, the pressure dependence of a transition involving the M point of the conduction band and the K point of the valence band[17] Now, to put numbers charateristics of hBN into equation (1), it is important to refer to the work of Solozhenko et al.[44] who directly measured the variation of the lattice parameters with P. One only requires the knowledge of the quantities $\left(\frac{\partial E_\alpha^i}{\partial a}\right)_c$ and $\left(\frac{\partial E_\alpha^i}{\partial c}\right)_a$ for both the fundamental conduction and valence bands. The data of Solozhenko et al.[44] have been fitted in term of pressure with a second order polynomial form , weighting the zero pressure value to a factor 5 relatively to each experimental value in the 1-12 GPa range. We obtain $c(P)/c_0 = 0.99671-2.126 \ 10^{-2} \ P + 7.18635 \ 10^{-4} \ P^2$ and $a(P)/a_0 = 1- 4.73413 \ 10^{-4} \ P + 6.49 \ 10^{-7} \ P^2$ , which indicates that the on-axis lattice parameter c is reduced by an amount of about 10 percent at 5 GPa whilst the reduction of the in-plane lattice a is about 45 times smaller at first order in P.

Therefore, neglecting variation of the dielectric constant with pressure, the variation of the fundamental bandgap under pressure $\frac{\partial E_g}{\partial P}$ can be reasonably approximated by its simplified version:

$$\frac{\partial E_g}{\partial P} \approx \ \delta E_M^C - \delta E_K^V \approx \ \left[\left(\frac{\partial E_M^C}{\partial c}\right)_a - \left(\frac{\partial E_K^V}{\partial c}\right)_a\right]\frac{dc}{dP} \qquad (5)$$

Further using for a simple calculation in the low pressure range, the value $\frac{dc}{dP}$ of Solozhenko et al.[44] and c= 0.66 nm,  we obtain $\frac{\partial E_g}{\partial c} \approx 2.15$ meVnm$^{-1}$.

Before to skip to the analysis of data we recorded for our levels emitting light in the 4 eV range, we have to allocate time to phonons which energies do not vary so dramatically under pressure, but that we need to discuss thanks to the indirect nature of the bandgap of hBN. Takashi Kuzuba et al.[45] have measured,  in the 0-11 GPa range, shifts of about 5 cm$^{-1}$/GPa and 4 cm$^{-1}$/GPa for the low frequency and high frequency E$_{2g}$ Raman active modes respectively. More recently, Saha et al. [46] have compared the behavior of Raman active phonons in bulk hBN and in BN nanotubes and they found respective slopes of 4.2 cm$^{-1}$/GPa and 4.3 cm$^{-1}$/GPa with a gradual broadening of linewidth at 0.5 cm$^{-1}$/GPa and 3 cm$^{-1}$/GPa,  respectively. The phase transition pressure to wurtzitic BN is about 13 GPa and the cubic phase is reached beyond 15 GPa. This detailed experiment is of value for us: it validates the lack of phase transition in the range of our experiments and indicates that the phonon broadening contributes to the broadening of our PL lines whilst it weakly contributes to the value of their pressure coefficient.

### III- 2  Pressure dependence of deep levels in hBN

The pressure dependence of the photoluminescence attributed to deep levels does not follow pressure shifts of extrema of the band structure. Under pressure all energy levels have a specific strategy, depending on their nature, symmetry, coupling with the band structure and with the phonon fields. The pressure dependence of these levels generally vary specifically with their depth in the bandgap as their wave function cannot be associated with a single valley due to competition of short range and long-range contribution to the potential of the defect[48-50]. Such competitions do vary from one defect to another[51]. Therefore, using simple words we can state that since the pressure coefficients of the photoluminescence energy at 4 eV in BN (~-25meV/GPa) measured through the series are smaller than the pressure coefficient of the bandgap of the bulk material (~ -36meV/GPa), it indicates that we deal with deep defects of hBN.

Recently, Xue et al.[10] studied the pressure coefficients of defects emitting in the 1.75 eV to 2.15 eV range in hBN flakes. These defects behave as SPS as the 4.1 eV defects were also found to operate[12]. However they exhibit smaller pressure coefficients, in general in the range between -10 to -6 meV/GPa, but they also measured a blue-shift for a line at 2.15 eV, at rate 14.1 meV/GPa). The energy shifts of the PL peaks vary with the emission energy as indicated in figure 3. Interestingly these authors also studied monolayers and bilayers and they found pressure coefficients of +1.3 meV/GPa and -11.9 meV/GPa. Based on the similarities between our shifts in the 3.3- 4 eV region and these ones in the 1.7 eV to 2 eV range, the positive pressure coefficient we measure for the 3.5 eV transition that rises up with increasing pressure is not an artefact and witnesses of the complex and composite nature of the emission with a superposition of the contributions of different actors in the 3.3 eV - 4 eV range in hBN. Unraveling new states upon application of pressure is not new. In GaAs, when increasing the pressure below values leading to the Γ- X crossover at a pressure of about 2 GPa, a new state rises from the conduction band and dominates the photoluminescence spectrum[52]. In hBN, the observation of positive coefficient for a state emitting light at ~3.5 eV reveals the existence of a recombination center very different from deep level centers. Finally, no phase transition was observed during our study, in line with the conditions of the experiment for pressures up to 5GPa, as expected.

## IV- CONCLUSION

We have studied the pressure dependence of the levels giving birth to the photoluminescence band in the 3.3-4 eV range in hBN. The pressure coefficient of most of the levels is negative, with a value of ~ -25 meV/GPa intermediate between the value measured for the indirect bandgap ~ -36 meV/GPa and the value measured for deep traps emitting near 2 eV (ranging between 0 and -15meV/GPa). We measure a positive pressure coefficient of +15 meV/GPa for a transition at ~ 3.4 eV and another peak observed at 3.35 eV which a pressure coefficient of -10 meV/GPa. Our investigation of the PL in the 3-4 eV region demonstrates, by using hydrostatic pressure, the composite nature of the photoluminescence in this region, with contributions of different actors. We hope this study will trigger new ones and will stimulate theoretical calculations.

**ACKNOWLEDGEMENTS**

This work and the PhD funding of C. Elias were financially supported by the network GaNeX (ANR-11- LABX-0014).

GaNeX belongs to the publicly funded Investissements d'Avenir program managed by the French ANR agency.

## FIGURE CAPTIONS

**Figure 1**: Low temperature photoluminescence (PL) spectrum of a piece of the BN crystal used to load the Diamond Anvil Cell. The photograph of the sample chosen with substantial enough PL in the 3.3 eV to 4 eV region to allow recording PL in the DAC at low T is inserted in the figure.

**Figure 2:** Pressure dependence of the low-temperature PL of deep defects in hexagonal boron nitride near 4 eV. Note the appearance of transitions at 3.4 eV, one of them red-shifts under increasing pressure (the low energy one) whilst its higher energy companion blue-shifts. All other levels redshift at approximatively the same rate of -25 meV/GPa

**Figure 3:** Pressure coefficients measured on BN bulk crystals and flakes:

(a) Blue spheres: this work, transitions in the 3.3-4 eV region, bulk crystal

(b) Red spheres: Ref. 10. transitions in the 2 eV region, bi-layer

(c) Gray spheres: Ref. 10. (figure 4) transitions in the 2 eV region, monolayer

(d) Green spheres: Ref. 10. (figure 3) transitions in the 2 eV region, monolayer

(e) Wine spheres: Ref. 42. absorption at 6 eV, bulk crystal

**Figure 1**

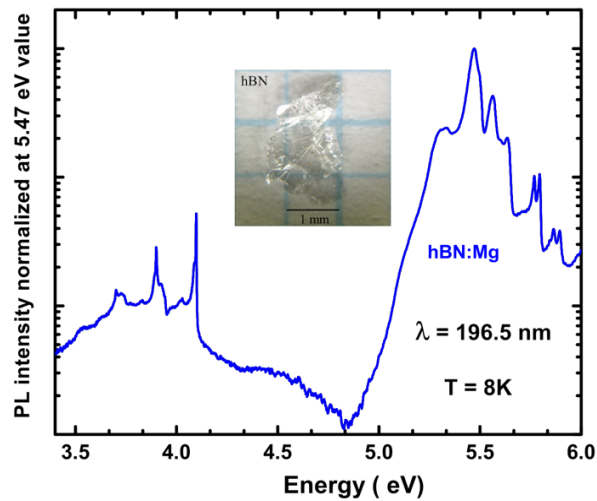

**Figure 2:**

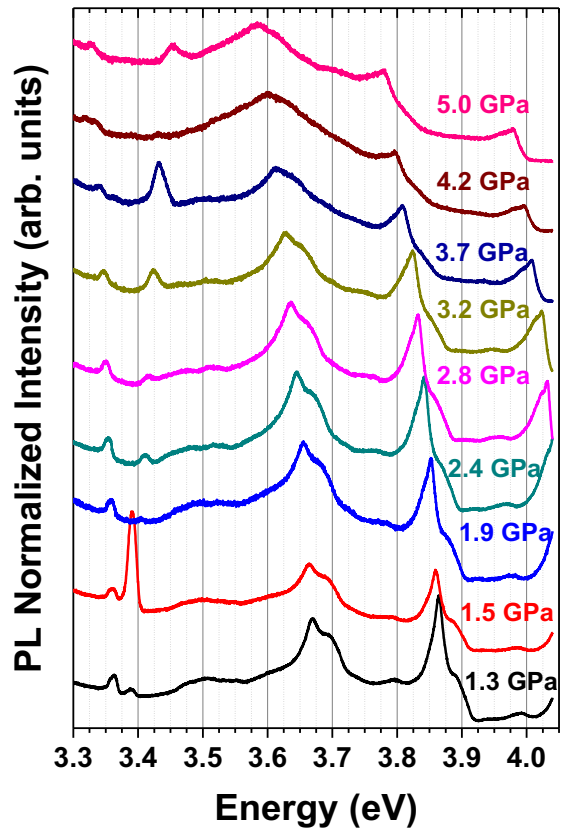

**Figure 3:**

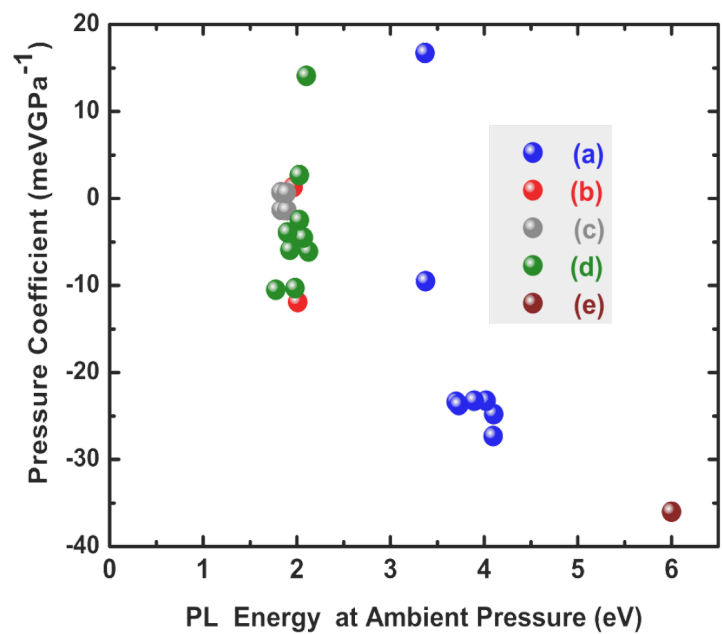